\documentclass[aps,prd,twocolumn,groupedaddress,amsmath,amssymb,showpacs]{revtex4}
\usepackage{graphicx}% Include figure files
\begin{document}

\title{Study of a formalism modeling \\massive particles at the speed of light \\on a Machian basis}

\author{Bur\c{c} G\"{o}kden}%read as Burch Goekden
\email{e119744@metu.edu.tr}
%\thanks{}
\affiliation{Middle East Technical University,\\ %Department of
%Electrical and Electronics Engineering,\\
06531 Ankara, Turkey}

\date{\today}

\begin{abstract}
In this paper we develop a formalism which models all massive
particles as travelling at the speed of light(c). This is done by
completing the 3-velocity $\mathbf{v}$ of a test particle to the
speed of light by adding an auxiliary 3-velocity component
$\mathbf{z}$ for the particle. According to the observations and
laws of physics defined in our spacetime these vectors are
generalized to domains and then two methods are developed to
define c domain in terms of our spacetime(v domain). By using
these methods the formalism is applied on relativistic quantum
theory and general theory of relativity. From these, the relation
between the formalism and Mach's principle is investigated. The
ideas and formulae developed from application of the formalism on
general relativity are compared with the characteristics of
anomalous accelerations detected on Pioneer 10/11, Ulysses and
Galileo spacecrafts and an explanation according to the formalism
is suggested. Possible relationships between Mach's principle and
the nondeterministic nature of the universe are also explored. In
this study Mach's principle, on which current debate still
continues, is expressed from an unconventional point of view, as a
naturally arising consequence of the formalism, and the approaches
are simplified accordingly.

\end{abstract}

%\keywords{}

\maketitle

\section{introduction}
It is a well known property of matter that a massive particle can
not reach the speed of light. Particles that travel at the speed
of light are observed to have no mass. Speed of light is a
constant of the universe and a particle travelling at the speed of
light is observed by all observers in spacetime travelling at this
constant speed. Inspired by this constancy of speed of light we
formed a theory which models all the matter in spacetime as matter
travelling at the speed of light so that no frame of reference in
the usual sense exists, all the particles have speed of light
relative to each other irrespective of directions of their motion.
Such a modelling satisfies and gives ideas and data encapsulating
Mach's principle, when it is applied on relativistic quantum
mechanics and general relativity; though no assumption in favor of
Mach's principle is made during the formation of the theory. For
this reason, before examining the details of the theory we will
present versions of Mach's principle as interpreted by quantum
theory and gravitation mainly referring to Feynman's discussions
on this subject \cite{Feynman}.

Mach's principle can be stated in different ways based on the
approach to the phenomenon to be described. Regarding the motion
of a particle, Mach's principle states that inertia represents the
effects of interactions with faraway matter. According to Mach the
concept of an absolute acceleration relative to absolute space was
not meaningful, the motion of a particle should be described
relative to distant matter in the universe. Similarly rotation
should be rotation relative to something else, not absolute
rotation. This definition of motion relative to distant matter may
have significant implications on the motion of a test particle,
since the usual mechanics assumes unaccelerated rectilinear motion
to be the natural motion in the absence of the forces. Feynman
noted in his Lectures on Gravitation \cite{Feynman} that ``when
accelerations are defined as accelerations relative to other
objects, the path of a particle under no acceleration depends on
the distributions of the other objects in space and the
definitions of forces between objects would be altered as we
change the distributions of other objects in space''. In a similar
manner he also argued that the absolute size of $g_{\mu\nu}$ which
is 1 in Lorentzian metric may as well be different than 1, flat
space may be $g_{00}=-g_{11}=-g_{22}=-g_{33}=\xi$ where $\xi$ is a
``meaningful number'' not to be simply taken as 1. Feynman's ideas
on Mach's principle show that Mach's principle may not only alter
the laws of mechanics but also gives constants of motion some
meaning.

On the quantum side, new fundamental units of length and time were
definable with the development of quantum mechanics. These were
the new absolutes of physics to be defined concerning Mach's
principle. These absolute units of time and length are the factors
that, for example, determine the maximum wavelength (or minimum
frequency) of a photon that can create electron--positron pairs.
Therefore, at each point of spacetime these absolute units must be
defined as a natural measure of size and time. From Mach's
viewpoint these natural measures are absolute only if they are
compared to something else; i.e., distant matter in the universe.
In other words, natural measures of size and time, such as Compton
wavelength or the term ${\hbar}/{mc^2}$ are influenced and
determined by distant matter in the universe. According to Mach
these units are not absolute if they are not specified relative to
something else.

Some results of Mach's principle stated above appear as a
consequence of making all matter travel at the speed of light in
the formalism which we develop in the next section.

\section{General Formalism}
\begin{figure}
\includegraphics{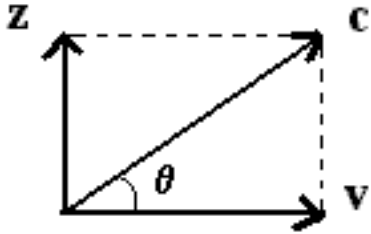}
\caption{\label{fig:fig1} visualization of the formalism as a sum
of two orthogonal vectors.}
\end{figure}

We set the speed of a massive test particle moving with velocity
$\mathbf{v}$ to the speed of light (c) by adding a nonoverlapping
vector $\mathbf{z}$ that completes $\mathbf{v}$ to c in magnitude
according to the simple expression
\begin{equation}
c^2 = \mathbf{v}^{2}+\mathbf{z}^{2} \label{eq:eq1}
\end{equation}
where $\mathbf{v}$ and $\mathbf{z}$ are 3-vectors in position
space. The geometric visualization of eq. (\ref{eq:eq1}) can be
seen as a simple diagram in fig. (\ref{fig:fig1}).

To explain why such a visualization is selected, we should define
constraints and properties loaded on these so-called vectors
$\mathbf{v}$, $\mathbf{z}$, and $\mathbf{c}$. To accomplish this
we let the characteristics of these vectors define their own
domains instead of defining all vectors in one position space,
where each domain has some properties compatible with observations
and characteristics of each vector they belong to. The domains
these vectors specify play a critical role in development of the
theory.

$\mathbf{v}$ vector defines the velocity of the test particle
measured in position space. One important property of the
$\mathbf{v}$ vector is its measurability, in other words it
specifies our local space as we observe in the vicinity of the
test particle. We assume that the laws of physics in this domain
are the same as those we define in our spacetime and neither
particles with mass can travel at the speed of light nor such a
massive object with speed c could be observed in this domain.
Since all measurements can be done only in this domain and current
laws of physics can not define equations of motion of a massive
object travelling at speed c, all effects of c domain that are
observable in the v domain are our only source of measurement and
we propose that \emph{an observer can observe the effects of any
domain on a test particle only in the v domain by using the laws
of physics defined in the v domain provided that any value (i.e.,
momentum, wave number) in that domain can be expressed in terms of
values defined in v domain.} We will call this proposition
\emph{method $\alpha$} for further reference. With this assumption
we say that if z domain or c domain has observable effects on the
particle and if these effects can be defined in terms of
parameters in v domain, then we can measure these effects in our
position space. But before using this assumption we should
identify z domain in terms of v domain and since v domain and z
domain form c domain, a relation defining c domain in terms of v
and z domains is needed.

Major property of z domain we assign to it is that it is an
unobservable domain and is totally unrelated from v domain so that
no effect observable in z domain is directly present in the v
domain. Due to this unrelatedness we represent $\mathbf{z}$ vector
as an orthogonal vector to the $\mathbf{v}$ vector forming the
simple diagram in fig. (\ref{fig:fig1}), which we use to develop
and understand the formalism in a geometrical way. Since we can
vary the parameters only in the v domain (i.e., $\mathbf{v}$
vector in the diagram), the range of z domain (i.e., range of its
possible effects, represented by the magnitude of $\mathbf{z}$
vector in the diagram) can only be changed by varying the
$\mathbf{v}$ vector while holding $\mathbf{c}$ constant in
magnitude. We do not specify any more properties for the z domain,
what entity it corresponds to in our space is an unknown to be
found by using our formalism on the phenomena we will inspect
later in this paper. But it is worth noting here the case that we
do not even specify z domain to be local to the test particle and
the $\mathbf{z}$ vector on the test particle even could be formed
according to a nonlocal procedure.

Since a massive test particle travelling at the speed of light is
not possible as observed in our spacetime(or equivalently in the v
domain) and also there is no physical law that defines the
dynamics of the massive particles at the speed of light, c domain
is an unobservable domain but it is a combination of both v domain
and z domain  as the diagram suggests. The c domain is expected to
include much more information than the v domain and z domain,
therefore, once  c domain is represented in terms of variables in
v domain (our only observable domain in the vicinity of the
particle) we will use the usual laws of physics to extract
information according to our assumptions stated as method
$\alpha$. In the diagram the $\mathbf{c}$ vector is defined as a
vector that has $\theta$ degrees between $\mathbf{v}$ vector and
$90^\circ-\theta$ degrees between $\mathbf{z}$ vector and this
helps us to specify the magnitudes of $\mathbf{v}$ and
$\mathbf{z}$ in terms of c, the speed of light. For the diagram we
should also note that we can vary the magnitude of $\mathbf{v}$
vector between 0 and c, and magnitude being equal to c corresponds
to the overlapping of the v domain with the c domain defining a
massless particle travelling at the speed of light in our space,
the only case c domain is totally observable and $\mathbf{z}=0$.
Thus $\theta$ has a range between $0^\circ$ and $90^\circ$,
$90^\circ$ corresponding to the case that the massive particle
stands still and $\mathbf{z}=\mathbf{c}$.

One last property which is common to all these domains is the
energy of the particle; in the c domain and z domain the energy of
a particle has the same value as the energy measured in the v
domain for that same particle. This should be valid, since no
extra energy is introduced while completing the speed of the
particle to the speed of light via addition of $\mathbf{z}$
vector.

The geometric diagram we presented in fig. (\ref{fig:fig1}) is
very useful as far as velocity of the test particle is taken into
consideration, but to generalize and develop the concepts of the
formalism we make an abstraction by defining domains. This will
enable us to propose a relation between the c domain and v domain
so that variables such as momentum and wave number in c domain are
definable in terms of variables in v domain, which in turn lets us
inquire the effects of the formalism via quantum theory and some
foundations of general relativity.

\subsection*{The relation between c domain and v domain}
In order to use method $\alpha$ it is needed to formulate a
relation between c domain and v domain so that c domain is
interpretable in terms of v domain, where our measurements and
laws of physics are guaranteed to be valid. To extract such a
formulation we will first examine our diagram.

According to the diagram the magnitudes of the $\mathbf{z}$ vector
and $\mathbf{v}$ vector are given as
\begin{eqnarray}
z&=&c\sin{\theta},\label{eq:eq2} \\
v&=&c\cos{\theta},~~~~~~~~~~0^\circ\leq\theta\leq 90^\circ
\label{eq:eq3}
\end{eqnarray}
where $z$ and $v$ are magnitudes of $\mathbf{z}$ and $\mathbf{v}$
vectors respectively. In v domain we can define the velocity of
the test particle at point $\mathcal{P}$ as
\begin{equation*}
\mathbf{v}=\frac{\partial{\mathcal{P}}}{\partial{t}}
\end{equation*}
where t is the time as measured by an observer in v domain. From
special relativity the 3-momentum of this test particle in v
domain is given as
\begin{equation}
\mathbf{p}=m\mathbf{u} \label{eq:eq4}
\end{equation}
where $\mathbf{u}$ is the spatial part of the four velocity of the
particle which can be defined in terms of spatial point
$\mathcal{P}$ and proper time $\tau$ of the test particle given as
\begin{equation}
\mathbf{u}=\frac{\partial{\mathcal{P}}}{\partial{\tau}}=\frac{\partial{\mathcal{P}}}{\partial{t}}\frac{dt}{d\tau}
=\mathbf{v}\frac{dt}{d\tau} \label{eq:eq5}
\end{equation}
where
\begin{equation}
\frac{dt}{d\tau}=\frac{1}{\sqrt{1-\dfrac{v^2}{c^2}}}
\label{eq:eq6}
\end{equation}
for an observer in an inertial reference frame. Combining these
results with the magnitude of $\mathbf{v}$ given in eq.
(\ref{eq:eq3}) we get
\begin{equation}
p=\frac{mv}{\sqrt{1-\dfrac{v^2}{c^2}}}=mc\cot{\theta}
\label{eq:eq7}
\end{equation}

If the observer in v domain in some way manages to define
$\mathbf{z}$ vector in terms of variables in v domain, we have
\begin{equation*}
\mathbf{z}=\frac{\partial{\mathcal{P'}}}{\partial{t}}
\end{equation*}
where $\mathcal{P'}$ is the spatial point in v domain that
corresponds to the effects of z domain. We can also define the
spatial part $\mathbf{w}$ of the test particle due to z domain
given as
\begin{equation}
\mathbf{w}=\frac{\partial{\mathcal{P'}}}{\partial{\tau}}
=\frac{\partial{\mathcal{P'}}}{\partial{t}}\frac{dt}{d\tau}
=\mathbf{z}\frac{dt}{d\tau} \label{eq:eq8}
\end{equation}
where $\frac{dt}{d\tau}$ is the same as eq. (\ref{eq:eq6}), since
all measurements are done in v domain according to method
$\alpha$. Implementing this result together with eq.
(\ref{eq:eq2}) into eq. (\ref{eq:eq4}) we have
\begin{equation}
p_z=mw=mc \label{eq:eq9}
\end{equation}

This means that whatever the magnitude of $\mathbf{z}$ is (even
when $z=0$ or $z=c$), the momentum associated with z domain
according to method $\alpha$ is a constant.

Although we were able to express momentum in terms of velocity for
z domain and v domain, we can not do such a momentum calculation
for c domain in terms of speed of light and the magnitude of
$\mathbf{v}$ vector as we did previously in eq. (\ref{eq:eq9}),
since from our observations in v domain, particles travelling at
the speed of light can have different values of momentum
independent of their speeds. We must not contradict with our
observations in the v domain if we want to construct a well
defined theory. For that reason, we should develop a new
formulation relating c domain to v domain so that we can apply
method $\alpha$.

One common property property of $\mathbf{v}$ vector and
$\mathbf{z}$ vector is that at all cases their magnitudes are less
than or equal to c. This enables c as being a constant in both
domains to be a base magnitude in calculations of relativistic
effects in terms of variables in v domain. Following this analogy,
we propose that if we can find some \emph{concept} which is not
bounded by c at all we could use this as a base to observe the
effects of c domain by using method $\alpha$. The wave-particle
duality of matter suggests a concept that satisfies this: the
phase velocity of a particle is always bigger than or equal to c
which is formulated as
\begin{equation}
s=\nu\lambda=\frac{c^2}{v} \label{eq:eq10}
\end{equation}
where $s$ is the phase velocity in v domain, $\nu$ is the
frequency given by $E/h$, E being energy of the particle and
$\lambda$ is the de Broglie wavelength of the particle given as
$h/p$. In terms of relativistic energy and momentum the phase
velocity is given as $c^2/v$ and since maximum value that $v$ can
have in v domain is c, the minimum phase velocity is c. By using
this property of matter waves and the general structure of the
formalism developed so far, we define the relation between the c
domain and v domain related to the same test particle through the
equation
\begin{equation}
s_c^2=s^2+s_z^2 \label{eq:eq11}
\end{equation}
where
\begin{eqnarray*}
s_c&=&\nu\lambda_{c}, \\
s_z&=&\nu\lambda_{z}, \\
s&=&\nu\lambda
\end{eqnarray*}
are the phase velocities in the c, z, and v domains respectively.

Since we assume that the energy of the particle is same in all
domains, eq. (\ref{eq:eq11}) can be simplified to
\begin{equation}
\lambda_{c}^2=\lambda^2+\lambda_{z}^2 \label{eq:eq12}
\end{equation}
Here $\lambda_{c}=h/p_c$, $\lambda_{z}=h/p_z$, $\lambda=h/p$ and
$p_c$, $p_z$, $p$ are magnitudes of momenta of the same particle
in c, z, and v domains respectively. It is clearly seen that eq.
(\ref{eq:eq12}) enables us to represent $p_c$ in terms of v domain
once we specify $p_z$ in terms of v domain. One thing we should
consider for eq. (\ref{eq:eq12}) is the boundary conditions which
must comply with our observations in the nature and the diagram
for a consistent theory. We know from the diagram that as $p$ goes
to 0, $\mathbf{v}$ goes to 0 in magnitude and z domain overlaps
with c domain which means that
\begin{equation*}
\lambda_{c}=\lambda_{z}
\end{equation*}
for $v=0$, $z=c$, and $p$ vanishes leading $\lambda$ to infinity.

For the case $p$ goes to infinity $v$ approaches to c and the
diagram suggests the overlapping of v and c domains. But according
to eq. (\ref{eq:eq12}) this is possible only if $p_z$ vanishes so
that $\lambda_{z}$ goes to infinity and $\lambda_{c}=\lambda$ is
satisfied. But from eq. (\ref{eq:eq9}) we showed that $p_z$ is a
constant nonzero value and for a massive test particle nonzero
$p_z$ even if $z=0$ means that it can never reach the speed of
light and c domain does not overlap with v domain completely.
However, foundation of our formalism offers that as $p$ goes to
infinity $v$ approaches very close to c and the diagram shows that
c domain almost overlaps v domain completely and from these
arguments we infer in the limit that
\begin{eqnarray}
&&\underset{p\rightarrow\infty}\lim{\lambda_{c}}\rightarrow\lambda
~\text{for a massive
particle,} \label{eq:eq13}\\
&&\lambda_{c}=\lambda ~\text{for a massless particle for any $p$.}
\label{eq:eq14}
\end{eqnarray}

With this result we mean that for a massive particle the values of
variables in c domain are very close to the values of variables
defined in v domain as $p$ goes to infinity so that we accept an
overlap in the limit. This result will be used when we consider
gravitation.

From the above discussion it is seen that constant $p_z$ at any
$z$ is the cause that prevents a massive particle from reaching
the speed of light. So this value defines a natural measure which
constrains the speed of a massive particle in v domain always to a
value under the speed of light and Mach's principle seems to
appear here. But to investigate if there is a rigid connection
between the formalism and Mach's principle, we need to identify
the effects of z domain better while examining c domain by using
method $\alpha$.

This completes the development of the formalism, in the next two
sections we will apply the formalism on the Dirac equation and
general theory of relativity  both to experiment it against well
known facts and to use it to explain phenomena that can not be
defined by current laws of physics.Through these sections we will
use eq. (\ref{eq:eq12}) in the arranged form
\begin{eqnarray}
p_c&=&\left(1+\frac{p^2}{p_z^2}\right)^{-\frac 1 2}p \label{eq:eq15} \\
k_c&=&\left(1+\frac{k^2}{k_z^2}\right)^{-\frac 1 2}k
\label{eq:eq16}
\end{eqnarray}
where $p_c=\hbar k_c$, $p_z=\hbar k_z$, $p=\hbar k$, and $k_c$,
$k_z$, $k$ are the wave numbers of the particle in c, z, and v
domains respectively.

We will refer to the concept eq. (\ref{eq:eq12}) stems from so
frequently that we call it \emph{method $\beta$} for simplicity.

\section{Application of the formalism to the Dirac equation}
The Dirac equation is the relativistic version of the
Schr\"{o}dinger's equation where the hamiltonian is the linearized
form of the relativistic energy--momentum equation given as
\begin{equation}
H=\sqrt{c^{2}P^{2}+m^{2}c^{4}}=c\,\boldsymbol{\alpha}\cdot\mathbf{P}+\beta
mc^2 \label{eq:eq17}
\end{equation}

Here the 4$\times$4 Dirac matrices $\boldsymbol{\alpha}$ and
$\beta$ are given in terms of Pauli and identity matrices as
\begin{equation}
\boldsymbol{\alpha}=
\begin{pmatrix}
0&\boldsymbol{\sigma} \\
\boldsymbol{\sigma}&0
\end{pmatrix}
~~\text{and}~~ \beta=
\begin{pmatrix}
I&0 \\
0&-I
\end{pmatrix}
\label{eq:eq18}
\end{equation}
with
\begin{equation*}
\boldsymbol{\sigma}=\sigma_1\mathbf{i}+\sigma_2\mathbf{j}
+\sigma_3\mathbf{k}=\begin{pmatrix}
0&1\\1&0\end{pmatrix}\mathbf{i}+\begin{pmatrix}0&-i\\i&0\end{pmatrix}
\mathbf{j}+\begin{pmatrix}1&0\\0&-1\end{pmatrix}\mathbf{k}\end{equation*}
and
\begin{equation*}
I=\begin{pmatrix}1&0\\0&1\end{pmatrix}\end{equation*}

The Dirac equation results in a multicomponent wave function which
not only includes spin information but also defines ``negative
energy'' solutions. When the electromagnetic interaction of a
Dirac particle is taken into account, the Dirac equation
successfully reduces in the nonrelativistic limit to the Pauli
equation for an electron \cite{Bjorken} and the gyromagnetic ratio
$g$ automatically emerges as $g=2$ which is very close to the
correct value. Then we expect the formalism to be compatible with
these results in the nonrelativistic limit and predict $g$ as good
as or better than the Dirac equation if it really describes the
physical world in a true manner.

To check if the above conditions are satisfied by the formalism we
will do our calculations in the nonrelativistic limit as well. For
that reason, method $\beta$ is used together with eq.
(\ref{eq:eq1}) to define c domain in terms of v domain for $v\ll
c$. When two sides of eq. (\ref{eq:eq1}) are multiplied by the
square of the mass, we have
\begin{equation}
m^2c^2=p^2+p_z^2 \label{eq:eq19}
\end{equation}
where we identify $p_z^2=m^2 z^2$ and $p^2=m^2 v^2$ in the limit
$v\ll c$. Thus, the relation between z domain and v domain is
satisfied and by inserting $p_z$ into eq. (\ref{eq:eq15}) we have
\begin{equation}
p_c=\left(\sqrt{1-\frac{p^2}{m^2 c^2}}~\right)p \label{eq:eq20}
\end{equation}

So far $p_c$ is considered in terms of magnitude of $\mathbf{p}$
but in applying the formalism on Dirac equation momentum operator
will be taken as a 3--dimensional operator, so an expression for
$\mathbf{p_c}$ whose magnitude satisfy eq. (\ref{eq:eq20}) is
required. The most general representation in terms of
corresponding momentum operators seems to be
\begin{equation}
\mathbf{P_c}=\left(\sqrt{1-\frac{\mathbf{P}^2}{m^2
c^2}}~\right)\mathbf{P} \label{eq:eq21}
\end{equation}
which does not prefer a predefined direction(as it might be
imposed by $\mathbf{v}$ vector according to the diagram but
generalization to domain concept helps us to use eq.
(\ref{eq:eq21}) with no restriction).

Before inserting $\mathbf{P_c}$ into the energy-momentum equation
and using the linearization procedure, we must linearize
$\mathbf{P_c}$ so that the equation we have in the end is local in
nature \cite{Bjorken}.For this reason the linearization first
takes place according to the equation
\begin{equation}
1-\frac{\mathbf{P}^2}{m^2
c^2}=\left(\gamma-\frac{\boldsymbol{\eta}\cdot\mathbf{P}}{mc}\right)^2
\label{eq:eq22} \end{equation}

By comparing both sides of this equation it can be found that the
following set of relations are satisfied by $\gamma$,
$\boldsymbol{\eta}$ where $\gamma$ and $\boldsymbol{\eta}$ are
taken as constant matrices.
\begin{eqnarray*}
&\gamma^2&=~~1, \\
&\left[\gamma, \eta_{i}\right]_+&=~~0,  \\
&\eta_{i}^2&=-1, \\
&\left[\eta_{i}, \eta_{j}\right]_+&=~~0,~~~~~~ i\neq j~~i,j=1,2,3
\end{eqnarray*}
where $\left[~~,~~\right]_+$ defines the anticommutator relation.

Matrices satisfying these relations can be given in terms of Dirac
matrices as
\begin{equation*}
\gamma=\beta\qquad\boldsymbol{\eta}=i\boldsymbol{\alpha}~.
\end{equation*}

Thus the linearized form of $\mathbf{P_c}$ is given as
\begin{equation}
\mathbf{P_c}=\left(\beta-i\frac{\boldsymbol{\alpha}\cdot\mathbf{P}}{mc}\right)\mathbf{P}~.
\label{eq:eq23}
\end{equation}

Having linearized $\mathbf{P_c}$, it is now possible to apply
method $\alpha$ by inserting $\mathbf{P_c}$ into the
energy--momentum equation. Performing the linearization process
for a second time according to the expression
\begin{equation}
H^2=\left(c^2\mathbf{P_c}^2+m^2
c^4\right)=\left(c\,\mathbf{C}\cdot\mathbf{P_c}+Dmc^2\right)^2
\label{eq:eq24}
\end{equation}
results in a large set of relations between $\mathbf{C}$, $D$,
$\gamma$ and $\boldsymbol{\eta}$. The matrices $\mathbf{C}$ and
$D$ obeying these relations are found to be
\begin{equation*}
\mathbf{C}=\begin{pmatrix}\boldsymbol{\sigma}&0\\0&\boldsymbol{\sigma}\end{pmatrix}
\qquad and \qquad D=i\begin{pmatrix}0&I\\-I&0\end{pmatrix}~.
\end{equation*}

With the derivation of $\mathbf{C}$ and $D$, all components of the
method $\alpha$ applied Dirac equation are found and the
electromagnetic interaction of an electron can be examined now.
The final form of the equation is
\begin{equation}
\left(c\,\mathbf{C}\cdot\mathbf{P_c}+Dmc^2\right)\Psi(\mathbf{x},t)=i\hbar\frac{\partial}{\partial
t}\Psi(\mathbf{x},t) \label{eq:eq25}
\end{equation}

To introduce the coupling of an external electromagnetic field to
an electron the momentum operator in v domain is substituted by
the momentum operator $\boldsymbol{\pi}$ given as
\begin{equation}
\boldsymbol{\pi}=\mathbf{P}-\frac{e}{c}\mathbf{A}~.
\label{eq:eq26}
\end{equation}

Here $\mathbf{A}$ is 3-vector potential of the electromagnetic
field. This $\boldsymbol{\pi}$ is substituted into $\mathbf{P_c}$
instead of $\mathbf{P}$ in the eq. (\ref{eq:eq21}).We also take
the scalar potential $\phi$ of the field as $\phi=0$ and as
already mentioned, work to order $\frac{v^2}{c^2}$ to see the
emerging of electron spin and gyromagnetic ratio from eq.
(\ref{eq:eq25}). We will look for energy eigenstates given as
\begin{equation}
\Psi(\mathbf{x}, t)=\Psi e^{-i{Et}/\hbar} \label{eq:eq27}
\end{equation}
where $E$ is the eigenvalue of the Hamiltonian. The sizes of
matrices $\mathbf{C}$ and $D$ implies $\Psi$ as a four component
wave function and it can be grouped into two component spinors.
This $\Psi$ in terms of two component spinors can be written as
\begin{equation}
\Psi=\begin{pmatrix}\chi\\\Phi\end{pmatrix} \label{eq:eq28}
\end{equation}

where $\chi$ and $\Phi$ are two component spinors. By inserting
eqs. (\ref{eq:eq26}), (\ref{eq:eq27}) and (\ref{eq:eq28}) into eq.
(\ref{eq:eq25}) as specified and after some modification of the
terms, eq. (\ref{eq:eq25}) becomes
\begin{equation}
\begin{bmatrix}
E-c\,\boldsymbol{\sigma}\cdot\boldsymbol{\pi}&&i\frac{\boldsymbol{\pi}^2}
{m}-imc^2\\i\frac{\boldsymbol{\pi}^2}{m}+imc^2&&E+c\,\boldsymbol{\sigma}\cdot\boldsymbol{\pi}
\end{bmatrix}\begin{pmatrix}\chi\\\Phi\end{pmatrix}=0
\label{eq:eq29} \end{equation}

By performing the matrix multiplication in eq. (\ref{eq:eq29}) we
have a set of two equations which are
\begin{subequations}
\label{eq:eq30}
\begin{equation}
\left(E-c\,\boldsymbol{\sigma}\cdot\boldsymbol{\pi}\right)\chi +
i\left(\frac{\boldsymbol{\pi}^2}{m}-mc^2\right)\Phi=0
\label{eq:eq30a}
\end{equation}
\begin{equation}
i\left(\frac{\boldsymbol{\pi}^2}{m}+mc^2\right)\chi +
\left(E+c\,\boldsymbol{\sigma}\cdot\boldsymbol{\pi}\right)\Phi=0
\label{eq:eq30b} \end{equation}
\end{subequations}

From eq. (\ref{eq:eq30b})
\begin{equation}
\chi=i\frac{1}{mc^2\left(1+\dfrac{\boldsymbol{\pi}^2}{m^2c^2}\right)}
\left(E+c\,\boldsymbol{\sigma}\cdot\boldsymbol{\pi}\right)\Phi
\label{eq:eq31}
\end{equation}

Here an operator in the denominator exists. This situation can be
grasped as follows. For two operators $A$ and $B$ we have the
relation \cite{Kaku}:
\begin{eqnarray}
\frac{1}{A+B} &=& A^{-1}-\frac{1}{A+B}BA^{-1} \nonumber \\ &=&
\left(1-A^{-1}B+A^{-1}BA^{-1}B+\cdots\right)A^{-1} \label{eq:eq32}
\end{eqnarray}

In our case this relation can be interpreted as an expansion of
Green's functions where $A^{-1}$ is the Green's function of free
particle ($G_0$) and $B$ corresponds to the small interacting
potential in terms of propagator theory(namely $H_I$). In terms of
Green's functions eq. (\ref{eq:eq32}) can be written as
\begin{equation}
G=G_0+GH_I G_0 \label{eq:eq33}
\end{equation}
where
\begin{eqnarray}
G_0&=&A^{-1}=\frac {1} {mc^2} \nonumber\\
H_I&=&-B=-\frac {\boldsymbol{\pi}^2} {m} \nonumber \\
G&=&\frac{1}{\left(mc^2+\dfrac{\boldsymbol{\pi}^2}{m}\right)} \nonumber \\
&=&\frac {1} {mc^2}\left(1-\frac{\boldsymbol{\pi}^2}{m^2 c^2} +
\left(-\frac{\boldsymbol{\pi}^2}{m^2 c^2}\right)^2+\cdots\right)
\label{eq:eq34} \end{eqnarray}

Note that with natural emerging of such a propagator--like
structure, the effects of virtual particles interacting with the
particle itself are taken into account in the wave equation we
develop; it means that the formalism introduces the basics of the
propagator theory into Dirac equation. This would exactly help us
to get information which was disjointly expressed before, by
examining one equation. The reason starting from eq.
(\ref{eq:eq30b}) is to see the effects of the total energy(i.e.,
$mc^2+ \frac{\boldsymbol{\pi}^2}{m}$) as a propagator--like
structure operating on the state $\Phi$ which seems to represent
the negative energy part of the Dirac solution.

To get a full equation, eq. (\ref{eq:eq34}) is inserted into eq.
(\ref{eq:eq31}) and eq. (\ref{eq:eq31}) is inserted into eq.
(\ref{eq:eq30a}) resulting in
\begin{equation}
i\left(\frac{\boldsymbol{\pi}^2}{m}-mc^2\right)\Phi +
i\left(E-c\boldsymbol{\sigma}\cdot\boldsymbol{\pi}\right)
G\left(E+c\boldsymbol{\sigma}\cdot\boldsymbol{\pi}\right)\Phi=0
\label{eq:eq35}
\end{equation}

Evaluating the terms and multiplying both sides by $1/2$ gives
\begin{widetext}
\begin{equation}
\biggl(\frac 1 2 EGE+\frac 1 2\,
cEG(\boldsymbol{\sigma}\cdot\boldsymbol{\pi}) - \frac 1 2\,
c(\boldsymbol{\sigma}\cdot\boldsymbol{\pi})GE - \frac 1 2\, c^2
(\boldsymbol{\sigma}\cdot\boldsymbol{\pi})G(\boldsymbol{\sigma}\cdot\boldsymbol{\pi})
- \frac 1 2
mc^2\left(1-\frac{\boldsymbol{\pi}^2}{m^2c^2}\right)\biggr)\Phi=0
\label{eq:eq36} \end{equation} \end{widetext}

As seen above $G$ handles all the situations: interaction of the
total energy of the particle with the total energy itself,
interaction of the total energy of the particle with the coupling
electromagnetic field, interaction of the coupling electromagnetic
field with the total energy of the particle and interaction of the
coupling electromagnetic field with the field itself respectively.
Since we work to order $\frac{v^2}{c^2}$ we take
\begin{eqnarray}
G&=&\frac{1}{mc^2}\left(1-\dfrac{\boldsymbol{\pi}^2}{m^2c^2}\right)
\label{eq:eq37} \\
E&=&mc^2+E_s \label{eq:eq38}
\end{eqnarray}
where $E_s$ is the energy eigenvalue that appears in the
Schr\"{o}dinger equation.

We then insert eqs. (\ref{eq:eq37}), (\ref{eq:eq38}) into eq.
(\ref{eq:eq36}) and after the cancellations are done we omit the
terms that are of higher order than $\frac{v^2}{c^2}$(these are
the terms including $E_s^2$,
$E_s(\boldsymbol{\sigma}\cdot\boldsymbol{\pi})$,
$\boldsymbol{\pi}^2(\boldsymbol{\sigma}\cdot\boldsymbol{\pi})$
etc.) and the resulting expression is
\begin{equation}
\left(\frac{\boldsymbol{\pi}^2}{2m} -
\frac{e\hbar}{2mc}\,\boldsymbol{\sigma}\cdot\mathbf{B}\right)\Phi=0
\label{eq:eq39}
\end{equation}
where the identities
\begin{eqnarray*}
(\boldsymbol{\sigma}\cdot\boldsymbol{\pi})
(\boldsymbol{\sigma}\cdot\boldsymbol{\pi}) &=&
\boldsymbol{\pi}\cdot\boldsymbol{\pi} +
i\boldsymbol{\sigma}\cdot(\boldsymbol{\pi} \times\boldsymbol{\pi})
\\ \boldsymbol{\pi}\times\boldsymbol{\pi} &=&
i\frac{e\hbar}{c}\mathbf{B}
\end{eqnarray*}
are used and $\mathbf{B}$ is the magnetic field.

It is clear that this equation is the Pauli equation and defines a
particle with spin $\frac 1 2$ and gyromagnetic ratio $g=2$. The
same equation can be also derived for the $\chi$ component, and
thus, in the classical approximation up to order $\frac{v^2}{c^2}$
the formalism is consistent with known facts. It can be also seen
from eq. (\ref{eq:eq36}) that expansion of $G$ to higher orders
suggests that gyromagnetic ratio slightly differs from 2, which is
due to a sum of powers of $\frac{\boldsymbol{\pi}}{mc}$, similar
to approximation of the gyromagnetic ratio as the sum of powers of
$\alpha\cong\frac 1 {137}$, the fine structure constant. This is
compatible with experiment and the propagator theory since an
electron can be as well presented in terms of one Dirac electron,
a Dirac electron and a photon, a Dirac electron and several
photons or several electron positron pairs etc. One thing we can
extract from eq. (\ref{eq:eq36}) is that this equation also
implies that spin is an intrinsic property of the particle. To
show this we take $G=\frac 1 {mc^2}=G_0$ and put this into eq.
(\ref{eq:eq36}). Taking $G=G_0$ means that no coupling has yet
resulted in motion of the particle itself(i.e., no motion is
observed in v domain). But it is a possibility that its
surroundings (or virtual particles, maybe) affecting the state of
the particle observed in v domain may interact with the field just
before the particle does. In other words we do not constrain a
simultaneity in the interacting times of v domain and z domain; it
is probable that z domain can interact before the v domain (recall
that we even do not underestimate the possibility of z domain
being non-local in nature and operators in squareroots might as
well lead to a nonlocal interpretation). If this is the case we
simply insert $G_0$ into eq. (\ref{eq:eq36}) and after omitting
the orders higher than $\frac {v^2}{c^2}$, it is seen that all
other terms exactly cancel and we have for $\Phi$(our z domain
correspondent)
\begin{equation}
\left(-\frac{e\hbar}{2mc}\,\boldsymbol{\sigma}\cdot\mathbf{B}\right)\Phi
= E_s\Phi \label{eq:eq40}
\end{equation}

Thus eq. (\ref{eq:eq40}) shows that the interaction hamiltonian
due to spin is active even if no momentum is present implying that
spin is an intrinsic property of the particle. Moreover, it also
shows that z domain introduces some measures and constants of size
and time as Mach's principle suggests(This last result is very
similar to the situation in eq. (\ref{eq:eq9})). Another finding
that can be inferred is that we can define the contents of z
domain in terms of the virtual particles and interactions in the
vicinity of the particle are taken into account, since an
interaction just before the particle interacts with the field may
result spin as an intrinsic property. But the definition of z
domain should be a general one, true for all interactions so we
move to gravitation in order to use the formalism to generalize
the ideas presented so far, since both Mach's principle and
spacetime have their most severe effects on gravitating bodies.

\section{From metric to Mach's principle}
Now by using the assumptions and results established so far, we
can develop a simple metric of spacetime to investigate the
effects of the formalism according to general theory of
relativity. The $g_{00}$ component of such a metric can be
developed by using method $\beta$ and method $\alpha$. By using
the concepts of method $\beta$ the eq. (\ref{eq:eq12}) can be
modified to represent the c domain only in terms of v domain where
the $g_{00}$ component of the metric relating v domain to the c
domain according to an observer in v domain can be given as
\begin{equation}
s_c^2 dt^2=g_{00}s^2 dt^2 \label{eq:eq41}
\end{equation}
Here v domain is taken as the base frame and the effects of z
domain are embedded into the metric tensor $g_{00}$ so that c
domain is interpreted as a curved spacetime relative to the v
domain. Here one novel thing is that we have used the phase
velocity in the metric implied by the eq. (\ref{eq:eq12}) and in
general by method $\beta$. And the measured time is the same in
both domains for that same test particle and any resulting effects
are loaded on the different values of the phase velocities. This
enables us to calculate $g_{00}$ in terms of phase velocities
\begin{equation}
g_{00}=\frac{s_c^2}{s^2} \label{eq:eq42}
\end{equation}

But if the effects of this metric are observable in v domain, it
should be observable according to the laws of physics and
variables observed in v domain. We know that differences of phase
velocities are not observed to cause relativistic effects in v
domain, but time dilation is a well known effect of the spacetime
depending on the gravitational potential at a point. Thus if such
a metric in eq. (\ref{eq:eq41}) has observable effects, it should
be observed as some type of time dilation and by taking the phase
velocity in v domain as a base, eq. (\ref{eq:eq41}) can be
restated as
\begin{equation}
s^2 d\tau^2=g_{00} s^2 dt^2 \label{eq:eq43}
\end{equation}
where $\tau$ is the proper time of the test particle that is due
to the c domain. Since we speak of the c domain and the v domain
associated with the same test particle we do not include spatial
terms in this metric in order to examine only the relations
between c and v domains. This metric does not mean that c domain
curves the spacetime really but it rather causes effects on the
test particle which seems to the observer as if extra
gravitational forces are applied on it.

Some properties of $g_{00}$ can be further investigated by using
eq. (\ref{eq:eq42}). Given the phase velocities
\begin{equation*}
s=\frac{\omega} {k}\qquad\qquad s_c=\frac{\omega}{k_c}
\end{equation*}
where $\omega$ is the angular frequency of the matter waves and
since we assumed the energy has the same value in all domains, it
is same in all of the expressions for a test particle. By
inserting these into eq. (\ref{eq:eq42}) and from eq.
(\ref{eq:eq16}) we have
\begin{equation}
g_{00}=\frac{s_c^2}{s^2}=\frac{k^2}{k_c^2}=1+\frac{k^2}{k_z^2}
\label{eq:eq44}
\end{equation}
So this shows that $g_{00}$ \emph{is a function of $k$ only} and
it does not depend on a single point in position space, and it is
a constant in spacetime as long as $k$ is constant.

Einstein stated in one of his communications with de Sitter
\cite{Barbour} that, for complete relativization of inertia the
metric should satisfy a set of boundary conditions, one of which
is $g_{00}(\mathbf{x})\rightarrow\infty^2$ as
$|\mathbf{x}|\rightarrow\infty$. The $g_{00}(\mathbf{k})$
component defined in eq. (\ref{eq:eq44}) obeys this boundary
condition as $|\mathbf{k}|\rightarrow\infty$ and the methods
defined in the formalism introduces Machian ideas into general
relativity.

According to the formulation of general relativity, an
approximation to the Newtonian gravitational potential can be made
in position space simply by using the metric of spacetime
approaching the Minkowski metric at infinity (with
$\eta_{00}(\mathbf{x})=1$ at $|\mathbf{x}|\rightarrow\infty$) in a
static universe. In a static universe we define the spacetime
metric by
\begin{equation}
c^2 d\tau^2=\eta_{00}(\mathbf{x}) c^2 dt^2 + |d\mathbf{x}|^2
\label{eq:eq45}
\end{equation}

From this metric the energy of a photon at a point $\mathbf{x}$ of
the space can be given as below by defining the proper frequency
$\nu_p$ as \cite{Frankel}
\begin{equation}
\nu_p=\frac{\nu}{\sqrt{\eta_{00}(\mathbf{x})}} \label{eq:eq46}
\end{equation}
and the energy is given by
\begin{equation}
\varepsilon(\mathbf{x})=\frac{h\nu}{\sqrt{\eta_{00}(\mathbf{x})}}
\label{eq:eq47} \end{equation} where
\begin{equation*}
\underset{|\,\mathbf{x}|\rightarrow\infty}\lim\eta_{00}(\mathbf{x})=1
\end{equation*}
So the gravitational potential $\phi$ at $\mathbf{x}$ is given by
\begin{equation}
\phi(\mathbf{x})=\frac{\varepsilon(\mathbf{x})-\varepsilon(\infty)}{\varepsilon(\mathbf{x})}
= 1-\sqrt{\eta_{00}(\mathbf{x})} \label{eq:eq48}
\end{equation}
which specifies an approximation to Newtonian gravitational
potential for metric approaching Minkowski metric as
$|\mathbf{x}|\rightarrow\infty$.

Similarly, in momentum space(namely in $\mathbf{k}$ space) a
potential can be defined by using the definition of potential in
position space. The energy we have according to eq.
(\ref{eq:eq43}) is
\begin{equation}
\varepsilon(\mathbf{k})=\frac{\hbar\omega}{\sqrt{g_{00}(\mathbf{k})}}
\label{eq:eq49}
\end{equation}

Now the value of $\varepsilon(\infty)$ should be found. In the
expression in eq. (\ref{eq:eq49}) $\omega$ also goes to infinity
as $|\mathbf{k}|\rightarrow\infty$ and this limit should be
compatible with the discussion resulting in the eqs.
(\ref{eq:eq13}) and (\ref{eq:eq14}). Therefore, in the limit
$|\mathbf{k}|\rightarrow\infty$ the variables in c domain have
values very close to the corresponding variables in v domain so
that c domain almost overlaps with c domain resulting in
\begin{equation}
\underset{|\,\mathbf{k}|\rightarrow\infty}\lim\varepsilon
(\mathbf{k})=\hbar\omega \label{eq:eq50}
\end{equation}

By using this we can define the potential in $\mathbf{k}$ space as
\begin{equation}
U(\mathbf{k})=\frac{\varepsilon(\mathbf{k})-\varepsilon(\infty)}
{\varepsilon(\mathbf{k})}=1-\sqrt{g_{00}(\mathbf{k})}
\label{eq:eq51}
\end{equation}

Now we must define to what physical entity this potential
corresponds. This value was derived by inserting the phase
velocity into the metric by using the analogy in general
relativity. What we have at hand is a potential depending on the
wave number of a test particle and; since this potential depends
on the wave number only, it is independent of changes in position
space. So we should seek data or experiment results suspecting or
detecting an effect of this kind, hoping that this helps us giving
a meaning to this potential in $\mathbf{k}$ space. One candidate
is the measurements of anomalous acceleration detected on the
Pioneer 10/11, Ulysses and Galileo spacecrafts \cite{Pioneer}.
Especially the data on Pioneers are very exact. The anomalous
accelerations of the Pioneers are measured as  $\sim
8\times10^{-10}\,m/s^2$ directed toward the Sun and these values
are measured to be fairly constant for a long time. Similar
effects are measured for the Ulysses and Galileo spacecrafts with
values of $\sim (12\times10^{-10}\pm3)\,m/s^2$ and
$\sim(8\times10^{-10}\pm3)\, m/s^2$ respectively and they are also
directed toward the Sun. The calculations Anderson et al.
presented show that the data for Ulysses and Galileo are highly
correlated with the solar radiation. But they also presented that
it is still possible in principle  for Ulysses to separate the
solar radiation effects from the anomalous acceleration whereas
for Galileo the measured acceleration does not seem to be reliable
so we will only present the anomalous acceleration the formalism
suggests for Galileo.

For velocities $v\ll c$, $\frac
{k^2}{k_z^2}\approx\frac{v^2}{c^2}$ and Pioneer 10 is measured to
have a constant speed of $12.2\times10^3\,m/s$ and Pioneer 11 is
measured to have a similar velocity. Inserting this into eq.
(\ref{eq:eq51}) by using the definition of $g_{00}$ in eq.
(\ref{eq:eq44}) the potential gives a value
\begin{equation}
U(\mathbf{k})=-8.27\times10^{-10} \label{eq:eq52}
\end{equation}
which is very close to the values measured for Pioneers. In a
similar manner we can calculate average speed of Ulysses which
followed an elliptical orbit during the measurement of anomalous
acceleration done by Anderson et al. By inserting the anomalous
acceleration into eq. (\ref{eq:eq51}) and solving for $v$ gives
\begin{equation}
v=14.696\times10^3\,m/s~. \label{eq:eq53}
\end{equation}

This value is only $194\,m/s$ higher than the average value of the
heliocentric speed measurements taken periodically during its
voyage from 5.4 AU in February 1992 near Jupiter to perihelion at
1.3 AU in February 1995 \cite{Ulyssesdata}.

And for the Galileo which has an approximately constant speed of
$7.19\times10^3\,m/s$ the $U(\mathbf{k})$ is calculated to be
$-2.8\times10^{-10}$. Compared to the value
$(8\pm3)\times10^{-10}$, this value is even smaller than the error
part and due to high correlation of this value with solar
radiation pressure, data from Galileo does not seem to be
reliable.

From the above discussion it is seen that the potential
$U(\mathbf{k})$ corresponds to the acceleration of a particle as a
physical variable. The characteristics of this acceleration is
that it is defined in terms of $\mathbf{k}$ and independent of the
position. Also the potential referring to an acceleration
magnitude does not explicitly specify a direction for the
acceleration but $g_{00}(\mathbf{k})$ being always greater than 1
suggests that it is always negative (i.e., opposes motion).

The success of potential $U(\mathbf{k})$ in defining the
magnitudes of anomalous accelerations of Pioneers and Ulysses
suggests that this anomalous acceleration is due to Mach's
principle which can be stated for this case as inertial properties
of objects are determined according to distant matter in the
universe. For the Pioneers' trajectories, which are heading out of
the Solar system and far enough from any strongly gravitating
system, the inertia of the spacecrafts respond differently to
their motion by opposing it due to Mach's principle. Ulysses'
trajectory, which was an elliptical orbit during the measurement
of the anomalous acceleration, also admits the intuition that its
inertia responds to the elliptical motion slightly differently and
this results as an anomalous acceleration that can not be
interpreted in terms of other gravitating bodies close to the
spacecraft. In other words, while the gravitational interactions
make Ulysses travel in an elliptical path, the resulting
elliptical motion makes inertia of Ulysses respond differently.
This interpretation may suggest inequivalence of inertial and
gravitational masses, but rather than accounting this anomaly to
the difference of inertial and gravitational masses we account
this to a direct effect of distant(in fact all) matter in the
universe, depending only on the motion of the particle according
to eq. (\ref{eq:eq51}). We are led to assume in this way because
there is an apparent distinction between these spacecrafts and
other celestial bodies such as planets in the universe. During
their journey, these spacecrafts do maneuvers and even change
their spin independent of the pure gravitational attraction
between the spacecrafts and the Solar system. And though Ulysses
is in an elliptical orbit, its motion is not completely governed
by the Sun and other planets, whereas the other celestial bodies
orbiting around the Sun have orbits determined by Sun and other
planets, which make these orbiting celestial bodies bound to some
conditions mainly determined by gravitation. But Ulysses' orbit is
not a bound orbit totally, the spacecraft can maneuver when needed
independent of the Sun's interaction with the spacecraft, making
it impossible to assign boundaries to all its motion only in terms
of its gravitational interaction with the Sun or other celestial
bodies. These differences could result in significant inertial
effects and the only boundary that can be used in modelling the
motion of such a particle may be the size of the universe or even
infinity. In fact our potential $U(\mathbf{k})$ helps us to answer
how all the matter in the universe can affect an object. This is
due to the fact that if $U(\mathbf{k})$ is a function of
$\mathbf{k}$ only then it is independent of position space and if
this $U(\mathbf{k})$ defines a constant value which we can measure
in position space (which is the case), we should be able to define
a relation between $U(\mathbf{k})$ and the position space. One
obvious connection between $U(\mathbf{k})$ and position space can
be the sum of fourier transforms of the potentials in the position
space, given as
\begin{equation}
U(\mathbf{k})=\frac{1}{2\pi}
\int^{+\infty}_{-\infty}\phi(\mathbf{x})\, e^{-i\mathbf{k}
\cdot\mathbf{x}}\, d^3\mathbf{x} \label{eq:eq54}
\end{equation}
where
\begin{equation*}
\phi(\mathbf{x})=\phi_{1}(\mathbf{x}) + \phi_{2}(\mathbf{x}) +
\phi_{3}(\mathbf{x}) + \cdots
\end{equation*}
is the sum of potentials and energy per unit mass terms that
affect the motion of the test particle at a point $\mathbf{x}$,
resulting from all matter and energy in the position space.

Eq. (\ref{eq:eq54}) defines $U(\mathbf{k})$ as superposition of
potentials in the position space which directly refers to the
statement that the matter in the universe affects the inertia of
the object, namely Mach's principle. This equation also explains
why $U(\mathbf{k})$ is constant since for any $\mathbf{k}$ all the
contributions of the potentials in position space are included so
that nothing is left in position space to change this
$U(\mathbf{k})$.

We have defined a value in $\mathbf{k}$ space as superposition of
its corresponding values in position space. We can backtrack what
we have done in assuming eq. (\ref{eq:eq54}), we can replace
$\mathbf{k}$ in $U(\mathbf{k})$ with $\mathbf{x}$ on the left hand
side and defining a potential as $U(\mathbf{x})$ means removing
the superposition of the potentials on the right hand side, which
results in
\begin{equation}
U(\mathbf{x})=\phi_1 (\mathbf{x}) + \phi_2 (\mathbf{x}) + \phi_3
(\mathbf{x}) + \cdots \label{eq:eq541}
\end{equation}
where superposition is eliminated by removing the fourier
transform.

Here the terms that have sources close to the particle will
survive so that the effects of the faraway matter will be
negligible as it is the case in position space.

Implications above arise completely from the relationships between
momentum and position spaces. In studies investigating the effects
of distant matter where position space is considered only, the
effects of distant matter show up by expressions implying
variations in values of constants of the universe. In this way,
the effects are valid in every point of space and independent of
position. In our formalism, when applied to general relativity,
the effects(as we take them as effects of distant matter)
including any change in values of basic constants arise due to the
relationships between the position space and momentum space whose
variables change independent of position, instead of a direct
change in values of constants of the universe. Sciama
\cite{Sciama} showed in his study on the origin of inertia that
the scalar gravitational potential of the whole universe satisfies
the condition
\begin{equation*}
G\Phi = -\,c^2
\end{equation*}
where $\Phi=-2\pi c^2 \rho\,\tau^2$ and $\rho$ is the mean density
of the matter in space and $\tau$ is a Hubble law related
constant. This equation relates the value of the gravitational
constant $G$ to the gravitational potential of the universe. This
result in fact is a consequence of many general relativistic
models of the universe and Feynman \cite{Feynman} uses this
expression to calculate the effects of distant matter on the term
$g_{00}$ in the vicinity of a star with mass $m$ giving (with
$\hbar=c=1$)
\begin{equation*}
g_{00}=1+\frac {2Gm} {r}
\end{equation*}
This expression is what our formalism suggests when $\frac
{k^2}{k^2_z} \approx \frac {v^2} {c^2}$ and $v^2=\frac {2Gm}{r}$.
It must be again pointed out here that the formalism applied to
general relativity gives the effect (acceleration) not the source
(potential) when the tensor element $g_{00}$ is depending on
momentum space variables, where we were expecting a potential
value in fact. Therefore this effect can be defined as fictitious
and source is defined as the distant matter (a nonlocalized
entity) according to the relationships between momentum and
position space. From point of view of the formalism, this
situation can be explained as a source in c domain which is
affecting v domain where we can not detect the source as a
localized point in v domain since c domain and thus the source
have an unobservable part in it, which is z domain. Such an
approach of the formalism also introduces inspection of the
variations and uncertainties in scales of measurements performed
on observables of a system in terms of the relationships between
momentum and position spaces.

Turning back to $U(\mathbf{k})$ again, we see another important
situation when the sum of the potentials $\phi(\mathbf{x})$ on a
test particle at a point $\mathbf{x}$ is constant. In this case we
have
\begin{equation}
U(\mathbf{k})=\phi\,\delta(\mathbf{k}) \label{eq:eq549}
\end{equation}
where $\delta(\mathbf{k})$ is the Dirac delta function and $\phi$
is a constant value.

Therefore, $U(\mathbf{k})$ is zero unless $\mathbf{k}=0$ for
$\phi(\mathbf{x})$ being constant. $\phi(\mathbf{x})$ constant
means that energy is either \emph{conserved} or it does not
\emph{depend} on $\mathbf{x}$ at all. For gravitating bodies
orbiting around another gravitating body, the total energy of the
system is conserved and $U(\mathbf{k})=0$ is satisfied, the
anomalous accelerations observed for Pioneers will not be observed
for orbiting bodies. This gives an explanation for not detecting
the same effects for the planets in the Solar system, which should
have been observed till this time if they were present. For
quantized systems where energy is quantized and does not depend on
position, the total energy is also constant and we again have
$U(\mathbf{k})=0$ unless $\mathbf{k}=0$. Having nonzero values
only for $\mathbf{k}=0$ further introduces the concept of
uncertainty in quantum mechanics, what this last result might mean
in terms of quantum theory is discussed below.

Since we established such a relation between $\mathbf{k}$ space,
position space and Mach's principle, we can use principles of
quantum mechanics to define further characteristics of the
universe according to the formalism.

From quantum mechanics it is a postulate that to every observable
there is an operator which is hermitian. Therefore, we can define
eq. (\ref{eq:eq51}) as an operator and since we can linearize it
as done for a similar expression in the previous section, we see
that this operator is in fact linearized in terms of momentum
operator. We can also define eq. (\ref{eq:eq51}) as a nonlocalized
operator by power expanding it in powers of momentum
operator(taking $\mathbf{P_z}$ as constant):
\begin{eqnarray*}
1-\sqrt{g_{00}(\mathbf{P})}&=&1-\sqrt{1+\frac{\mathbf{P}^2}{\mathbf{P_z}^2}} \\
&=& 1-\left(1+\frac 1 2 \frac{\mathbf{P}^2}{\mathbf{P_z}^2} -
\frac 1 8 \frac{\mathbf{P}^4}{\mathbf{P_z}^4} + \cdots\right)
\end{eqnarray*}
Quantum mechanics also states that for each operator, its
observables are the eigenvalues of that operator which must be
real. It is a well known issue that for every free particle state
we can define an eigenvalue which is $\hbar\mathbf{k}$ for
momentum operator but for bound particles there are no eigenvalues
for the momentum operator satisfying the boundary conditions
\cite{Nettel}.This situation, during the development of the
quantum theory in its early times, made the first significant
difference of the quantum theory from classical theory that the
expected value should be used and determinism was lost. Since the
operator $1-\sqrt{g_{00}(\mathbf{P})}$ is also dependent on the
momentum operator, there is also no eigenvalue for this operator
satisfying the boundary conditions for a bound particle. In terms
of the diagram in fig. (\ref{fig:fig1}), assigning boundary
conditions means limiting the allowed values in v domain, this in
turn means limiting the z domain which introduced the effects of
the whole universe. z domain being limited only to some
permissible values means that one can not introduce all the
effects of the matter in the universe since boundary conditions in
v domain allow only some selected part of the points in the
universe to interact with the bounded system. This means that
$U(\mathbf{k})$ is no more constant and does not cover the effects
of all matter in the universe for that system, the excluded part
of the universe may as well affect $U(\mathbf{k})$ and change the
constant measures of size and time it specifies. This situation
can be also interpreted in such a way that for a bounded system
each point in the universe defines its own constants of measures
of size and time so that measurements taken at different points in
spacetime \emph{measures the events according to different scales}
(e.g., each point in spacetime may define different Compton
wavelengths). This means that determinism is lost, we have no
absolutes of time and size same for all points in the universe
valid for all time instants and we can not fully predict future
dynamics of a system depending on its past dynamics.

This situation could as well be generalized to include all other
bound objects. These bound objects also constrain the v domain and
z domain and this makes the constantness of $U(\mathbf{k})$ be
destroyed since there is still matter that does not satisfy the
bounds both in v domain and in z domain and only the expected
values are measured as a result.

This is an expected characteristic of our universe indeed,
elementary particles and fields seem to consist of quanta and
these quanta form bounded systems which are also quantized due to
their building blocks and for such a system one can not make the
whole universe to satisfy the boundary conditions of the bounded
system so that we can not define constant $U(\mathbf{k})$ or other
measures of size and time valid for all points. So this also
answers the questions why $U(\mathbf{k})$ is observable in our
space if the observed particle is free and why we don't say that
equivalence principle is violated for the Ulysses or Pioneer case.
In fact equivalence principle applies only to experiments that are
isolated from the rest of the universe, otherwise it results in
paradoxes \cite{Barbour}. Moreover, these discussions suggest an
answer to the question what entity z domain corresponds to, it
seems that z domain corresponds to the rest of the universe(or
position space in general) that interacts with the test particle.

One more implication of the formalism is due to the phase velocity
we used to establish these ideas. If we accept that the effects
presented in this study are propagating with the phase velocity
which is always larger than $c$ for matter waves, this enables us
to define Lorentz frames where a particle can move on straight
lines free of any gravitational effects of other bodies or stand
still, and in all these cases Mach's principle stating that
inertial frames are determined relative to distant stars can be
satisfied. For distant matter the phase velocity carries the
effects of inertia faster than light speed and before the
causality between a free test particle and distant matter occurs,
the inertial properties of the mass can be already transmitted.
Thus, until the causality is established(we assume that gravitons
also travel at the speed of light) the free particle is in a
Lorentz frame where all inertial effects are seen. Note that since
$g_{00}(\mathbf{k})$ is independent of position space we can
further assume that the Lorentz frame has $g_{00}=1$ when
$\mathbf{k}=0$ and $g_{00}(\mathbf{k})>1$ for free particle in
motion.

\section{conclusions}
In this paper we developed a formalism which models all particles
travelling at the speed of light and then applied it on two
distinct concepts, namely the quantum theory and general
relativity. In doing these, it has been seen that the concept of
phase velocity and wave-particle duality of matter played a
critical role.

One important implication of the formalism is that it includes the
Machian viewpoint, and embeds Mach's principle in other theories
via the contribution of z domain into our space, when it is
applied on these theories. For general relativity, the formalism
results in ideas that were developed by several physicists
including Einstein, but the formalism approaches these ideas
differently. Einstein spent much effort to make his theory of
general relativity Machian \cite{Barbour}, he proposed to develop
a Machian theory by imposing boundary conditions to get rid of
non-Machian solutions of general relativity and following these
ideas he was led to the idea of a \emph{closed} universe to make
the theory Machian. And Einstein's theory has the concept of
Minkowski metric as a valid solution which is totally non-Machian
in character.

The formalism faces these problems as well, boundary conditions
became an important case that should be taken care of. The
formalism handles the situation as a special case of the core
principle stating that matter here is governed by matter there,
such that the scales determined by distant matter for the bounded
system are not nonchanging scales valid at every point of the four
dimensional manifold and only a selected set of points in space
time have a common scale, but rest of the points in spacetime
define their own scale leading to measurements being statistical
in nature. Therefore, Mach's principle is not violated, matter
still depends on other matter in spacetime, but this relativeness
is no more a strict distinction between the matter and the rest of
the universe; the matter could also have the concept of
relativeness defined differently by different points in the
universe leading a way to the so-called loss of determinism. Once
this is accepted, there is no need to impose conditions such as
being closed and finite for the universe from a Machian viewpoint.
Moreover, the inclusion of phase velocity in the formalism also
enables a procedure for the construction of a Minkowski metric
where all inertial effects due to the distant matter are
established but no gravitational effect of that distant matter has
been applied yet.

%\begin{acknowledgments}
%I am grateful to G\"{o}zde Bozda\u{g}i Akar for her constant
%encouragement and to Sibel Ba\c{s}kal for her support.
%\end{acknowledgments}

%\bibliography{theformalism}

\end{document}